\documentclass[doublecol]{epl2}
\usepackage{graphicx}
\usepackage{dcolumn}
\usepackage{bm}
\usepackage[colorlinks,linkcolor=blue,citecolor=blue,dvipdfm]{hyperref}
\usepackage{multirow} 
\usepackage{booktabs,threeparttable} 

\begin{document}
\title{Transversal Propagation of Helical Edge States in Quantum Spin Hall Systems}

\author{Feng Lu\inst{1}, Yuan Zhou\inst{1}, Jin An\inst{1} and Chang-De Gong\inst{2,1}}

\institute{
    \inst{1} National Laboratory of Solid State Microstructure, Department of Physics, Nanjing University - Nanjing 210093, China \\
    \inst{2} Center for Statistical and Theoretical Condensed Matter Physics, Zhejiang Normal University - Jinhua 321004, China}


\abstract{The transversal propagation of the edge states in a two-dimensional quantum spin Hall (QSH) system
is classified by the characteristic parameter $\lambda$. There are two different types of the helical
edge states, the normal and special edge states, exhibiting distinct behaviors. The penetration depth
of the normal edge state is momentum dependent, and the finite gap for edge band decays monotonously with
sample width, leading to the normal finite size effect. In contrast, the penetration depth maintains a uniform
minimal value in the special edge states, and consequently the finite gap decays non-monotonously
with sample width, leading to the anomalous finite size effect. To demonstrate their difference explicitly,
we compared the real materials in phase diagram. An intuitive way to search for the special edge
states in the two-dimensional QSH system is also proposed.}

\date{\today}
\maketitle

\section{Introduction}
Owing to the linear dispersion and $Z_{2}$ topological
invariant, the anomalous transport properties of the helical edge states in a
quantum spin Hall (QSH)system are predicted\cite{km:Z2-05,helicaledge,BHZ:model,km:graphene-05,QSH:T&E-J,TSC}.
Well localized edge states were usually treated as ideal one-dimensional
channels to investigate the exotic properties\cite{helicaledge,QSH:T&E-J}.
However, their behaviors are significantly changed due to the transversal broadening of these edge states in real samples. The penetration depth of helical edge states had been discussed in both {\em HgTe} quantum well system and {\em Bi} thin
film\cite{QSH:pd-Bi}. In those two systems, its value are determined by the
inverse of momentum space distance between the edge state and the
absorption point into the bulk. The finite penetration depth also leads to the so-called finite size effect in
two-dimensional (2D) QSH system\cite{QSH:finite-size}. A gap opens
at $\Gamma$ point when the opposite helical edge channels overlap
each other, which had been used to confirm the intrinsic spin
Hall effect in {\em HgTe} quantum well
system\cite{QSH:finite_size_exp(1)}. Recently, several electric
devices had been designed to manipulate the charge and spin
transport with such
finite size effect\cite{finite-size-switch,QSH:spinmodulate}. An anomalous finite size effect was
further reported in three-dimensional ($3$D) topological insulator
$Bi_{2}Se_{3}$, much shorter penetration
depth and oscillatory finite size
gap had been revealed\cite{3Danomalous}. This gap oscillation had
also been used to search new candidate of topological non-trivial
systems\cite{OscalCross:2Dto3D,TIfilm:model-detail}. These previous
discussions were all based on the specific materials. The
general comprehension of the transversal propagation behaviors is
expected in QSH system, especially the intrinsic difference between the normal and anomalous finite size effects.

In this paper, the transversal propagation behaviors of the helical
edge states in 2D QSH system are investigated. The helical edge states can be classified into two modes, the normal and special edge states, according to the decay
characteristic quantity $\lambda$. In normal edge states, the penetration depth shows clear momentum-dependence, and the finite gap for edge states decays monotonously with the sample width. While in special edge states,  the penetration depth keeps unchanged in the momentum space, and its finite gap decays oscilatorily with sample width. The normal and anomalous finite size effect can be found in the respective edge states.
These facts give explicit explanations on the difference between the real materials. Based on the theoretical calculations, the search of the special edge states in the $2$D case is proposed.

The paper is organized as follows: In Sec.II, we
specify two different transversal propagation modes of the helical
edge states without specific boundary condition. A semi-finite
boundary condition is adopted to show the distinct evolution of the
penetration depth in the normal and special edge states in
Sec.III. As a consequent effect, the normal and
anomalous finite size effects are discussed in
Sec.IV. In Sec.V, the role of
particle-hole asymmetry and the comparison of real materials are
further discussed. The conclusion is drawn in Sec.VI.

\section{Transversal Modes of 2D QSH Model}
The QSH effect was theoretically predicted in $HgTe$
quantum well\cite{qshe,BHZ:model}, and soon confirmed experimentally
by K\"{o}nig {\em el at}.\cite{Kronig:qsh-qw}. We start from the
effective 4$\times$4 model for a 2D QSH system proposed by Bernevig,
Hughes, and Zhang\cite{QSH:T&E-J,BHZ:model}. Very recently, it
was also adopted as an effective 2D model for the 3D topological
insulator in ultrathin limit\cite{TIfilm:model}. The model
Hamiltonian is expressed as
\begin{equation}
\label{Eqn1}
H(k)=\left[\begin{array}{cc}
             h(k) & 0 \\
             0 & h^{*}(-k)
           \end{array}
\right],
\end{equation}
where
$h(k)=\varepsilon_{k}\mathbf{I}_{2\times2}+\mathbf{d}_{k}\cdot{\boldsymbol\sigma}$,
with $\varepsilon_{k}=C-D(k^{2}_{x}+k^{2}_{y})$. The k-dependent
effective field $\mathbf{d}_{k}=(Ak_{x},-Ak_{y},M_{k})$, where
$M_{k}=M-B(k^{2}_{x}+k^{2}_{y})$. $\boldsymbol\sigma$ is the Pauli matrices. $A$, $B$, $C$, $D$, and $M$ are determined by the quantum well geometry in real materials. Here, we treat
them as independent parameters and study their respective role
first. Keep in mind that, the interested topological non-trivial QSH
phase\cite{QSH-phase} emerges only when $MB>0$. In {\em HgTe}
quantum well system, such condition is controlled by the thickness
of the quantum well\cite{BHZ:model,Kronig:qsh-qw}. The properties of counter-part $h^{*}(-k)$ can be conveniently obtained by applying the time reversal operation to $h_{k}$.

To focus on the edge properties, $k_{y}$ needs to be replaced by
$-i\partial_{y}$, while $k_{x}$ remains a good quantum number due to the
translational symmetry. A trial solution of
$\Psi_{k_{x}}(y)=C_{k_{x}}e^{-\lambda y}$ can be introduced, and the
decay characteristic quantities are subsequently obtained as\cite{QSH:finite-size}
\begin{equation} \label{Eqn2}
\lambda_{1,2}^{2}(k_{x},E)=k_{x}^{2}+F(E)\pm\sqrt{F^{2}(E)+\frac{E^{2}-M^{2}}{B^{2}-D^{2}}},
\end{equation}
where $F(E)=\frac{A^{2}-2(MB+DE)}{2(B^{2}-D^{2})}$ is a function of energy $E$.

The transversal propagation behaviors of the states are determined
by both of $\lambda_{1,2}$. However, it is clear in
Eqs.~(\ref{Eqn2}) that, $\lambda(k_{x},E)$ has a definite
distribution in momentum space, independent of the boundary
condition. Hence, we can directly discuss the transversal
propagation behaviors of the states from Eqs.~(\ref{Eqn2}).
There are four modes of the
states specified by different combinations of $\lambda_{1,2}$, as
illustrated in TABLE~\ref{Tab1}.

\begin{threeparttable}[hbtp]
\caption{Combination for $\lambda_{1,2}$}
\label{Tab1}
\renewcommand{\arraystretch}{1.2}
\centering
\begin{tabular}{p{4 em}p{6 em}p{5 em}p{5 em}}
\hline
Mode\tnote{*} & Condition & $\lambda_{1}$ & $\lambda_{2}$ \\
\hline
Edge1 & $\lambda_{1}^{2}>\lambda_{2}^{2}\geq0$ & Real & Real \\
Bulk1\tnote{**} & $\lambda_{1}^{2}\geq0>\lambda_{2}^{2}$ & Real & Imaginary \\
Bulk2 & $0>\lambda_{1}^{2}>\lambda_{2}^{2}$ & Imaginary & Imaginary\\
Edge2 & $\lambda_{1}^{2}=(\lambda_{2}^{2})^{*}$ & Complex & Complex \\
\hline
\end{tabular}
\begin{tablenotes}
\footnotesize
\item[*] For bulk states, at least one of $\lambda_{1,2}$ is purely imaginary.
While for edge states, both of $\lambda_{1,2}$ should have a
non-zero real part.
\item[**] Trivial edge states are also included due to the real $\lambda_{1}$.
\end{tablenotes}
\end{threeparttable}
\bigskip

Since only the non-trivial edge states are interested in this paper, it is
natural to ask whether there is something different between the two
edge modes in TABLE~\ref{Tab1}. For convenience, we
specify the Edge1 state with both $\lambda_{1,2}$ real as the
normal edge state (NES); and the Edge2 state with $\lambda_{1,2}$
complex conjugates as the special edge state (SES). To fulfill the condition of conjugate, the term under square root in Eqs.~(\ref{Eqn2}) must be negative, which restricts SES existing in a specific regime in momentum space
confined by
\begin{equation}
\label{Eqn3}
E^{SES}_{\pm}=D\left(\frac{A^{2}-2MB}{2B^{2}}\right)\pm|\frac{A}{2B}|\sqrt{\gamma\left(4MB-A^{2}\right)}
\end{equation}
Here a screen factor denoting the particle-hole asymmetry
$\gamma=1-\frac{D^{2}}{B^{2}}$ is introduced. The system undergoes
a phase transition from an insulator to a
semimetal when $D\geq B$\cite{QSH:E-H-asym}, which is not interested for us.
Eqs.~(\ref{Eqn3}) naturally requires
$4MB\geq A^{2}\geq0$, implying a non-trivial QSH state. There
is no special restriction for the NES.

\begin{figure*}[t]
\onefigure[width=13.5cm]{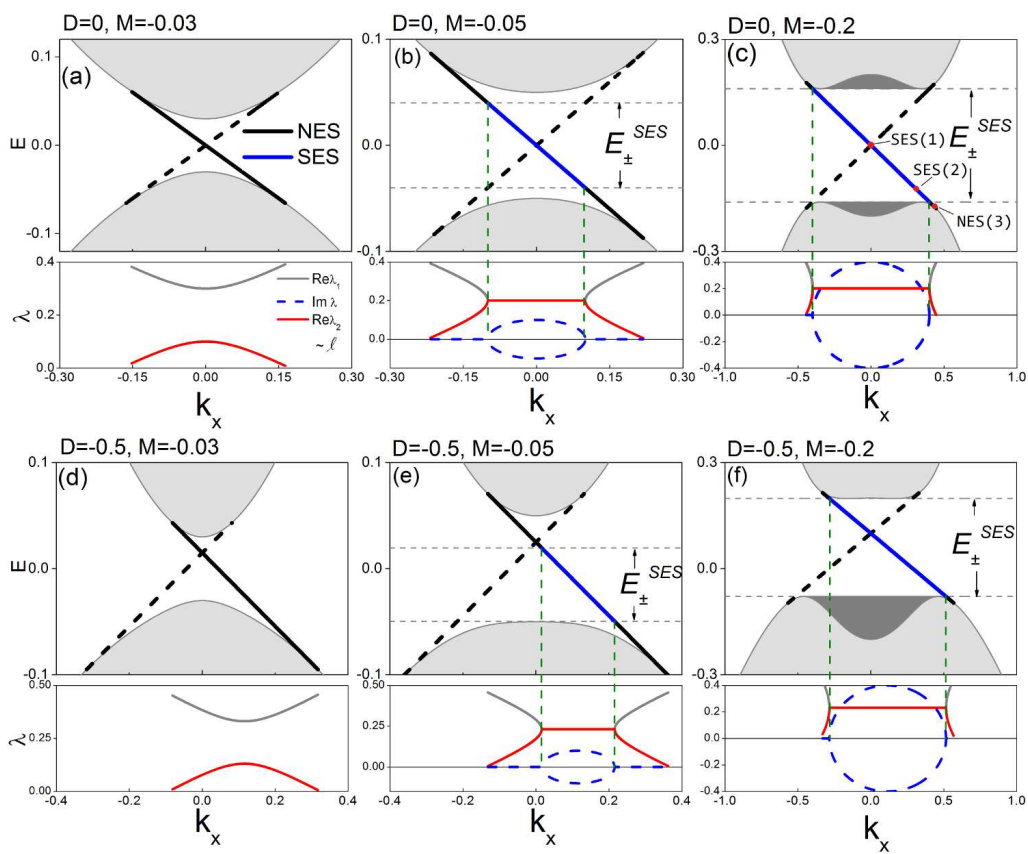}
\caption{Helical edge spectra for different effective parameters.
$A=0.4$ and $B=-1.0$. The upper/lower panels are spectra with/without
particle-hole symmetry. $|M|$ increases from left to right. The upper, and lower part in each panel is the edge spectrum, and corresponding $\lambda_{1,2}$ respectively.
The bold solid line in black/blue represents the edge spectrum of NES/SES. The lighter/darker gray regimes describe
the Bulk1/Bulk2 states in Table~\ref{Tab1}. The light grey dash line are the confines of SES given by Eqs.~(\ref{Eqn3}). The red solid lines in lower part of each panel is $\Re\lambda_{2}$, which give the inverse of $\ell$.
The blue dash lines are the corresponding $\Im\lambda_{1,2}$. In panel (c), three states are specified, and corresponding transversal propagation behaviors in real space are shown in Fig.~\ref{Fig2}.}
\label{Fig1}
\end{figure*}

Present classification is a natrual consequence of the breaking of periodic boundary condition, which leads to a definite distribution of transversal propagation modes in momentum space. An explicit boundary condition just creates a specific spectrum onto such distribution. The emergence of SES is determined by the $k_{x}$-independent $E_{\pm}^{SES}$. For given parameters, $E^{SES}_{\pm}$ will squish the bulk band, leading to a flat valence band top (or conduct band bottom), as in Fig.~\ref{Fig1}(c)(e)(f). Conversely, such feature can be viewed as a sufficient condition for SES in QSH system, even without the explicit knowledge of material parameters. Moreover, the squished bulk band also gives rise to the Bulk2 states in TABLE~\ref{Tab1}, which exhibit a larger density of states than Bulk1 states in our numerical results. Such classification is also significant in the problems of interference tunneling and restricted edge transport\cite{QSH:tunneling1,QSH:spinmodulate}.
More differences of the edge modes will be discussed in the following sections.

\section{Penetration Depth of Helical Edge States}
The penetration depth distribution of the helical edge states is
distinct from that of the chiral edge states in integer quantum
Hall effect\cite{chiral-ES(1),TKNN}. The latter is determined by
the universal magnetic length, which is related to the external
magnetic field. In contrast, the penetration depth in QSH system is $k$-dependent, originated from the band structure\cite{QSH:T&E-J}. To address this, the semi-infinite boundary
condition\cite{QSH:pd-Bi} is adopted here. We restrict
$\Re\lambda_{1,2}\geq0$ to obtain an evanescent edge state
localized near the boundary. The boundary condition
$\Psi_{k_{x}}(0)=0$ gives the linear dispersion
relation\cite{QSH:pd-Bi,QSH:finite-size}
\begin{equation}
\label{Eqn4}
E=-A\sqrt{\gamma}k_{x}-\frac{MD}{B}
\end{equation}

The penetration depth
$\ell=max\{\Re\lambda_{1,2}^{-1}\}$\cite{TSC} behaves
differently in NES and SES. The NES situation had been discussed
in previous work as in Fig.~\ref{Fig1}(a) and (d). The
$k_{x}$-dependent $\ell_{NES}$ reaches its minimum at
$k_{x}=\frac{A^{2}}{4}(1-\gamma^{-1})$. The edge state is absorbed
by the bulk when $\Re\lambda_{2}=0$. In contrast, the
penetration depth in SES maintains a uniform minimal value across the
whole regime of SES, given by
\begin{equation}
\label{Eqn5}
\ell_{SES}=2/(\lambda_{1}+\lambda_{2})=|\frac{2\sqrt{\gamma}B}{A}|,
\end{equation}
which is also independent of $E$, $k_{x}$ and $M$, as shown in Fig.~\ref{Fig1}. Here $\Re\lambda_{1,2}$ governs the transversal decay
behavior. In fact, although the relation
$\lambda_{1}+\lambda_{2}=\frac{A}{\sqrt{\gamma}B}$ keeps unchanged
even in NES, the penetration depth in NES is merely determined by the
minimum of real $\lambda_{1,2}$.

\begin{figure}[tbp]
\onefigure[width=8.5cm]{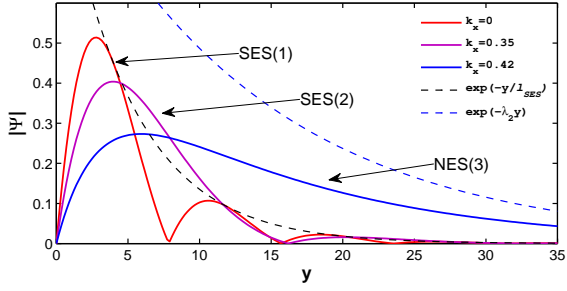}
\caption{Transversal propagation behaviors of edge state wave
function in real space. Corresponding states are marked in
Fig.~\ref{Fig1}(c). The dash line roughly gives the
penetration depth behaviors with function $exp(-y/\ell)$. The
black dash line takes $\ell_{SES}$ given by
Eqs.~(\ref{Eqn5}), while the blue one takes
$\ell=\Re\lambda_{2}^{-1}$.}
\label{Fig2}
\end{figure}

In Fig.~\ref{Fig2}, the transversal propagation behaviors of three selected helical edge states are plotted in real space. The wave function of SES(1) and SES(2) exhibit an evanescent oscillation with different periods. However, they share the same penetration depth. In contrast, the NES(3) shows no oscillation but much longer penetration depth. The non-monotonous decay behavior of edge state was also reported in the lattice model\cite{finite-size:lattice}, which can be naturally attributed to the SES.

The {\em HgTe} quantum well\cite{BHZ:model} and the ultra-thin $Bi_{2}Se_{3}$ film\cite{TIfilm:model} correspond to the situation in Fig.~\ref{Fig1}(d), where the SES is absent. The penetration depth is estimated to be about 50 nm\cite{QSH:finite-size}.
In previous studies,\cite{QSH:pd-Bi} the $Bi\{111\}$ thin film was compared with the {\em HgTe} quantum well system, and remarkable difference was found in the behaviors of the penetration depth. We notice that, a flat valence band top emerges in $Bi\{111\}$ spectrum\cite{QSH:pd-Bi}, implying the existence of SES. Therefore, such difference can be well understood within present discussion.
Due to the similarity of $\lambda$ at $\Gamma$ point, the topological surface states (TSS) of the 3D topological insulator can be equivalently discussed within our framework, corresponding to the situation in Fig.~\ref{Fig1}(f), where the edge states are SES dominated. The penetration depth of TSS in 3D $Bi_{2}Se_{3}$ was also reported in previous work, with a shorter $\ell$ of about 10 nm\cite{3Danomalous}. However, they concluded $\ell$ is proportional to the inverse of $\left\vert M\right\vert $, distinguished from present discussion. In fact, this situation does not belong to NES, but SES, since both the decay characteristic quantities $\lambda_{1,2}$ have image part as they stressed, too. Therefore, the penetration depth should be independent of $\left\vert M\right\vert $. The difference between NES and SES will be further discussed in the next section.

\section{Normal And Anomalous Finite Size Effects}
The finite size effect in QSH system arises from the overlap of the
opposite channel due to the decreasing sample width, leading to the finite energy gap opening for the energy dispersion of edge state near Dirac point\cite{QSH:finite-size}. Since the penetration depth of NES
and SES is quite different, the consequent finite size effect is
also expected to be distinct. We now turn to the ribbon geometry with the boundary condition of
$\Psi_{k_{x}}(-L/2)=\Psi_{k_{x}}(L/2)=0$, where $L$ is the width of the ribbon. Our numerical results reveal that, the relative gap
$\delta\Delta(k_{x})=\Delta(k_{x})-2|Ak_{x}|$ reaches its maximum at
$\Gamma$ point and decays exponentially with $|k_{x}|$. Hence, we
just focus on the situation at $\Gamma$ point where $\delta\Delta(0)=\Delta(0)$. We follow the previous discussions\cite{QSH:finite-size,TIfilm:model-detail} to evaluate the finite size gap in different situations.

\begin{figure}[tbp]
\onefigure[width=8cm]{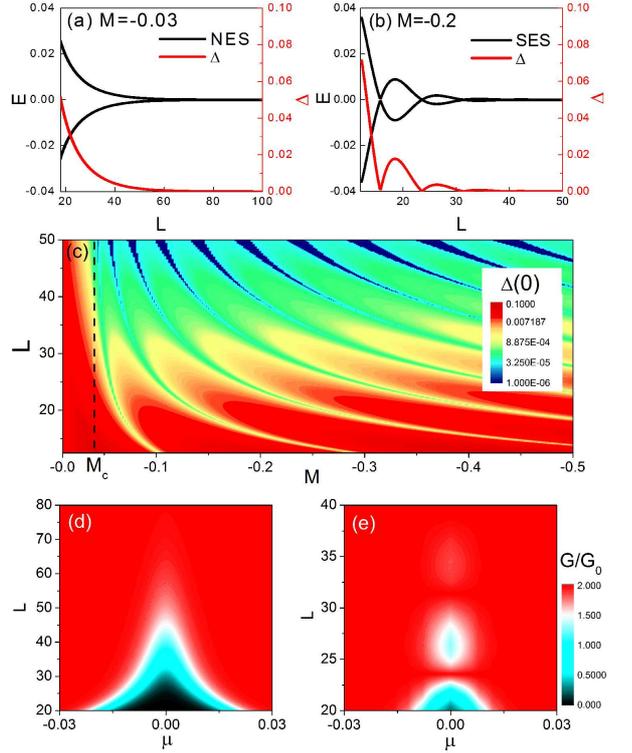}
\caption{Normal and anomalous finite size effects. (a), and (b) show
the edge bands and gap behaviors varying with $L$ in NES, and SES,
respectively. (c) gives the $M$-dependent $\Delta(0)$ with the same parameter adopted in Fig.~\ref{Fig1} (in logarithmic
scale). The normal (left), and anomalous (right) finite size effect
is divided by the critical $M_{c}$ (black dash line). (d), and (e)
present the conductance at finite temperature corresponding to the
situation of (a), and (b), respectively, with $1/k_{B}T=400$ and $G_{0}=\frac{e^{2}}{h}$.}
\label{Fig3}
\end{figure}

When the NES dominates the Dirac point, as discussed previously in
the {\em HgTe} quantum well\cite{QSH:finite-size}, the gap was
estimated to be
\begin{equation} \label{Eqn7} \Delta(0)\simeq
\frac{4|AM\gamma|}{\sqrt{A^{2}-4MB\gamma}}e^{-\lambda_{2}L}.
\end{equation}
Here we assume $\lambda_{1}L>>1$ and $\lambda_{1}>>\lambda_{2}$.
This is the normal finite size effect as shown in
Fig.~\ref{Fig3}(a). When Dirac point locates inside the SES
regime, the gap turns to be
\begin{equation}
\label{Eqn8}
\Delta(0)\simeq \frac{8|AM\gamma \sin(\Im\lambda_{2}L)|}{\sqrt{A^{2}-4MB\gamma}}e^{-\ell_{SES}^{-1}L},
\end{equation}
here $\Im\lambda_{2}=\sqrt{\frac{M}{B}-\frac{A^{2}}{4\gamma B^{2}}}$
is the imaginary part of $\lambda_{2}$. The gap exhibits an
oscillatory behavior with $L$, as described in
Fig.~\ref{Fig3}(b). This oscillation was also predicted in 3D topological
insulator\cite{3Danomalous,TIfilm:model-detail}, referred as the anomalous finite
size effect.

We numerically investigate the $M$-dependent evolution of $\Delta(0)$ to
distinguish the difference between NES and SES as shown in
Fig.~\ref{Fig3}(c). Here $D=0$ is applied to avoid the
mismatch between the Dirac point and the regime of SES. For small
$|M|$, the Dirac point is NES, and the corresponding $\Delta(0)$
evolves monotonously with $|M|$. For large $L$, the Dirac point
turns to be SES, $\Delta(0)$ is oscillatory. A critical
$|M_{c}|=0.04$ is obtained with the same parameters taken in
Fig.~\ref{Fig1}. The number of oscillatory periods increase
with $|M|$, owing to a decreasing $\Im\lambda$. Considerable gap
always opens at  $L\sim35$ (arb. units) for all $|M|>|M_{c}|$,
which coincides with the uniform minimal
$\ell_{SES}$ discussed above. Here we emphasize that, the uniform minimal $\ell_{SES}$ found in Fig.~\ref{Fig1} is protected by the linear dispersion of Eqs.~(\ref{Eqn4}). This linear relation is not preserved when the finite size gap opens, then $\ell$ becomes momentum-dependent again even in SES.

The essence of such differences can be understood based on present results. For SES, a $y$-dependent phase factor emerges due to the finite $\Im \lambda$, which
is absent in NES. The edge band is renormalized, together with gap
opening, due to the overlap of opposite edge states. Meanwhile, the
transversal phase coherence of the opposite edge states contributes
to the oscillation of $\Delta$ for SES. Hence, the interference-fringe-like picture can be obtained as shown in Fig.~\ref{Fig3}(c).

Such effect can be detected in transport measurements at low
temperature\cite{QSH:finite_size_exp(1)}. The conductance at finite temperature is simply given by\cite{QSH:finite-size}
\begin{equation}
G(\mu,T)=(2e^{2}/h)\left[f(\Delta/2-\mu)-f(-\Delta/2-\mu)+1\right]
\end{equation}
when $\mu$ locates inside the bulk gap. Here
$f(E)$ is the Fermi distribution function and $\Delta$ is the finite
size gap. Fig.~\ref{Fig3}(d) and Fig.~\ref{Fig3}(e)
present the conductance for
NES and SES respectively. Recently, a non-monotonous gap evolution
had been observed in ultra-thin $Bi_{2}Se_{3}$, which is
noted as a possible anomalous finite size effect for
TSS\cite{TIfilm:exp(2)}.

\section{Discussion}
\begin{figure}[tp]
\onefigure[width=8.5cm]{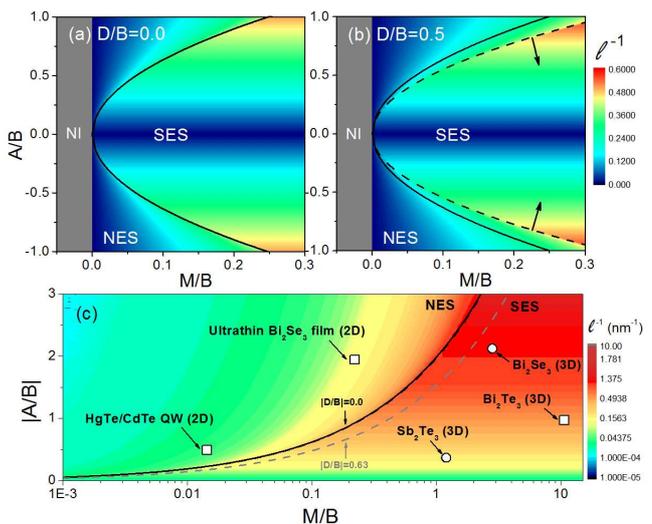}
\caption{Phase diagrams in parameter space. (a), and (b)
are with, and without the particle-hole symmetry respectively. The
black solid line ($A^{2}=4MB$) divides the region of edge state into SES and NES. The dash line ($A^{2}=4\gamma MB$) in (b) indicates the mismatch between SES and the Dirac point as described in the text, and moves along the direction indicated by arrow when $\gamma$ decreases. The intensity stands for the inverse of penetration depth $\ell^{-1}$ at Dirac point. Different
materials are compared in (c), where $A$, $B$ and
$M$ are unified into the units of $meV$ and $nm$, so
that $\ell$ has a common unit of $nm$. Here the logarithmic scale
is adopted. Materials with/without SES Dirac point are marked with cycles/squares. The intensity in (c) is for $\gamma=1$. For $Bi_{2}Se_{3}$ TSS, and $Sb_{2}Te_{3}$, $\vert D/B \vert$ is $0.13$, and $0.63$, respectively.}
\label{Fig4}
\end{figure}

Up to now, we have discussed the penetration depth and the finite
size effect in NES and SES. The particle-hole asymmetry factor
$\gamma$ also plays a subtle role on these properties. The Dirac point moves upward, and the SES regime also
shifts, leading to the possible mismatch as shown in
Fig.~\ref{Fig1}(e). The existence of SES Dirac point requires
$\frac{A^{2}}{B^{2}}\leq\gamma\frac{4M}{B}$. In
Fig.~\ref{Fig4}(b), the regime between the solid line and the
dash line describes the mismatch: the SES exists, but the Dirac
point moves outside. It should be pointed out that such mismatch is not sensitive with selected $|D/B|$ unless it approaches to $1$.

The penetration depth of the edge states at Dirac point and
finite size effect are discussed together in the phase space of
relative parameters $\frac{A}{B}$ and $\frac{M}{B}$, as shown in
Fig.~\ref{Fig4}. In these phase diagrams, since the solid line
divides the parameter space into two regimes, $\ell^{-l}$ at Dirac point reveals two
distinct evolutions. $\ell^{-1}$ at Dirac point increases with $M/B$ and decreases with $|A/B|$ in NES. In contrast, it remains unchanged with $M/B$ but increases with $|A/B|$ in SES.
The recently discovered topological non-trivial systems: 2D {\em HgTe} quantum
well\cite{QSH:finite-size}, Ultrathin $Bi_{2}Se_{3}$
film\cite{TIfilm:model}, 3D $Bi_{2}Se_{3}$, $Sb_{2}Te_{3}$ and $Bi_{2}Te_{3}$
\cite{3DTI:4-candidates,TSC} are compared in the same phase
space. The former two effective 2D systems are described by the
same model of Eqs.~(\ref{Eqn1}). Although the 3D
topological insulators have a different effective model\cite{3DTI:4-candidates}, the
situation at Dirac point is equivalent to the two 2D systems
under proper parameter substitution\cite{TIfilm:model-detail,3Danomalous}. As in
Fig.~\ref{Fig4}(c), the helical edge states in the two 2D
systems contain only NES, therefore, large size is required to
avoid the normal finite size effect. In contrast, the 2D TSS of 3D
$Bi_{2}Se_{3}$ and $Sb_{2}Te_{3}$ implies a shorter penetration depth and a possible
anomalous finite size effect\cite{3Danomalous,TIfilm:exp(2)}. Interestingly, the SES exits in $Bi_{2}Te_{3}$, however, its Dirac point moves into bulk state due to strong particle-hole asymmetry mentioned above. This may be true as compared with the angle resolved photoemission spectroscopy measurements\cite{adma:bi2te3,TSC}. Similar behavior may also can be found in the bulk $HgTe$ under uniaxial strain\cite{DX:HgTe}. We expect that, these special effects can be electrically detected
in other QSH systems with smaller $A/B$ or larger $M/B$ as shown in
Fig.~\ref{Fig4}. Recently, several designs had been performed based
on the finite size effect\cite{finite-size-switch,QSH:spinmodulate}. The future
applications could be quite sensitive to these properties. In this
sense, present work provides a theoretical prediction on the
possible finite size effects in new materials.

\section{Conclusion}
In conclusion, two different transversal propagation modes of the
helical edge states, i.e., NES and SES, in the QSH system are specified by the
decay characteristic quantities $\lambda$. The emergence of the
flat bulk band implies the special edge state, which gives a
sufficient criterion to distinguish the two modes. The penetration
depth of SES keeps a uniform minimal value, independent
of the selected $E$, $k_x$ and $M$. In
contrast, it is much larger and shows clear
momentum dependence in NES. Different finite size effects are
studied in respective edge states. Especially, the oscillatory gap for edge band
is found in SES. Some real materials are compared in the
phase diagram to demonstrate the difference between NES and SES. We also give clues to search possible QSH materials with
SES for future applications.

\acknowledgments
We would like to thank Y. F. Wang, and L. Xu for helpful
discussions. This work is supported by NSFC Project No. 10804047, and A Project Funded by the Priority Academic Program Development of Jiangsu Higher Education Institutions. J. An acknowledges NSFC Project
No. 10804073. C. D. Gong also acknowledges 973 Projects No. 2009CB929504.

\end{document}